\documentclass[12pt]{article}
\usepackage{graphicx}
\usepackage{float}
\title{Longitudinal Coherence Enhancement of X-Ray Free Electron Lasers}
\author{Paul J. Channell \\
paul.j.channell@gmail.com
\date{} }

\def\ie{i.e.\ }

\chardef\other=12
\def\ttverbatim{\begingroup \catcode`\\=\other
\catcode`\{=\other
\catcode`\}=\other \catcode`\$=\other
\catcode`\&=\other
\catcode`\#=\other \catcode`\%=\other
\catcode`\~=\other
\catcode`\_=\other \catcode`\^=\other
\obeyspaces \obeylines \tt}
{\obeyspaces\gdef {\ }}

\outer\def\begintt{$$\let\par=\endgraf \ttverbatim
\parskip=0pt
\catcode`\|=0 \rightskip=-5pc \ttfinish}
{\catcode`\|=0
|catcode`|\=\other
|obeylines
|gdef|ttfinish#1^^M#2\endtt{#1|vbox{#2}|endgroup$$}}

\catcode`\|=\active
{\obeylines\gdef|{\ttverbatim\spaceskip=\ttglue\let^^M=\
\let|=\endgroup}}

\begin{document}

\maketitle

\begin{abstract}
We present a new scheme for establishing longitudinal coherence
in the output of X-ray FELs. It uses a sequence of undulators and
delay chicanes and the careful tailoring of the initial e-beam with
time dependent focusing. Simulation of a simplified model
of the FEL shows that the idea works well.
\end{abstract}

\section{Introduction}

\begin{figure}[H]
\begin{center}
 \includegraphics[scale=0.40]{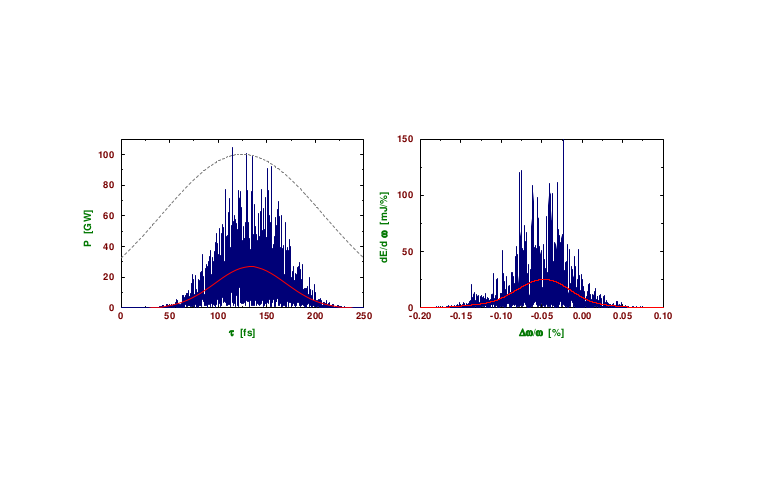} 
\end{center}
\vspace{-0.4in}
\caption{Typical longitudinal output of an X-ray FEL.}
\label{fig:spike}
\end{figure}

Though the new generation of X-ray free electron lasers (FELs) produces X-rays that are 
almost completely coherent transversally, longitudinally the output 
is `spiky', as indicated roughly in figure \ref{fig:spike}; \ie the spectral 
bandwith is larger than desired, \cite{survey}. 
\noindent Schemes have been 
proposed to deal with this problem such as the Geloni idea, 
\cite{mono1}, \cite{mono2}, and echo enhanced harmonic  
generation to produce seeds in the soft X-ray regime, \cite{stupakov},
\cite{hemsing}. 
In this paper we will outline a new approach to this problem that 
might extrapolate more easily to shorter wavelengths and be more 
easily tunable.

In the next section we discuss the basic idea of our approach,
noting that a set of undulators and e-beam chicanes 
combined with initially longitudinally decreasing wave 
intensity might be used to increase coherence lengths. 
In section $3$ we show that an  initial photon beam 
with decreasing intensity can be created by tailored 
electron beam size variations.  
In section $4$ we present simulations of a reduced model
of the X-ray FEL that indicate that our idea seems to work
well. In the final section we discuss our results
and point out further work with more realistic models
that is required.
In appendix A we present the Maxima code used in the simulations.
In appendix B we discuss a simple example of how to create the tailored 
e-beam size.

\section{Basic Idea}

Typically a spike covers a coherence length and is longitudinally 
coherent over it; similarly the other spikes are about a coherence 
length in duration and are longitudinally coherent but are not in 
phase with the other spikes. One approach might be to delay the 
electron beam by using a chicane consisting of dipoles
to increase the electron travel distance by a coherence length 
during the chicane. The photon bunch then travels forward in the
electron bunch by a coherence length (\ie the electrons fall 
behind by a coherence length), leaving electrons nearer the tail 
bunched in phase with the photons now one slice ahead but with very few photons. 
In the next section of undulator the electron oscillations toward the 
tail of the bunch rapidly grow at the same phase as the now forward 
photons since the electrons are strongly seeded. Of course, this 
won't increase the coherence length of the photons completely because 
the photons that are moved forward in the electron bunch will interfere 
with the phase of the electrons in the forward section and tend to cancel
the field. To overcome this cancellation we need to add one more idea.

(Note that a chicane consisting of a positive dipole followed
by a negative dipole is probably adequate, though a combination
of a half positive dipole, followed by a full negative dipole
followed by a half positive dipole probably is more 
achromatic and has smaller higher order terms.)

\begin{figure}[H]
\begin{center}
 \includegraphics[scale=0.30]{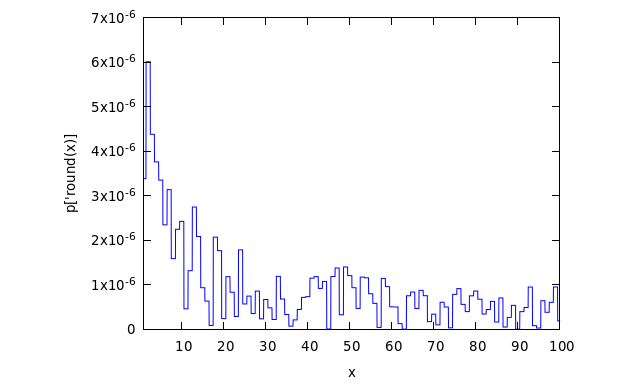} 
\end{center}
\caption{Initial decreasing longitudinal power with bunch length of an X-ray FEL.}
\label{fig:initdecrease}
\end{figure}

Suppose instead that the laser intensity, after a few growth times, 
decreases along the bunch, being stronger toward the rear and weaker
toward the front, as shown in figure \ref{fig:initdecrease}. 
\noindent In this case, when 
the photon spike moves forward  it encounters electrons with 
less bunching at the wrong phase and can convert the electron 
bunching to the right phase. After a few growth 
lengths, the coherence length toward the tail is longer. 
Repeating this process, the next time we  slide the photons 
forward we need to go through fewer growth lengths because 
of the larger amplitudes. After a few repetitions of this process, 
essentially the entire photon bunch should be longitudinally coherent.

\section{Decreasing Longitudinal Intensity}

Of course we have to show how to achieve the temporal output shown 
in figure \ref{fig:initdecrease}. To do this we refer to the growth formulae of 
\cite{huang-kim}, \cite{survey}. The FEL power at a point $z$ down the undulator 
is given by

\begin{equation}
P=P_0e^{z/L_G},
\end{equation}

\noindent where $P_0$ is the initial power and $L_G$ is the growth length. 
The growth length is given by

\begin{equation}
L_G={\Lambda_u \over 4\pi \sqrt{3} \rho},
\end{equation} 

\noindent where $\Lambda_u$ is the undulator period and 
$\rho$ is the Pierce parameter given by

\begin{equation}
\rho = \left[ {1\over 16}{I_e\over I_A} {K_0^2\left[ JJ\right] ^2 \over \gamma_0^3\sigma_x^2k_u^2}\right]^{1\over 3}, 
\end{equation}

\noindent with $I_e$ the beam current, $I_A$ the Alfven current, $K_0$ the 
undulator parameter, $k_u$ is the undulator wave number, $\gamma_0$ is 
the relativistic e-beam factor, $\left[ JJ\right]=J_0(\xi)-J_1(\xi) $, 
(Bessel functions), with
$\xi = K_0^2/(4+2K_0^2)$ (planar wiggler), and  $\sigma_x$ is the RMS 
transverse beam size. Note the dependence on beam
size, $\sigma_x$. We can rewrite the growth length as

\begin{equation}
L_G=\chi \sigma_x^{2/3},
\end{equation}

\noindent where $\chi$ contains all the remaining factors of $L_G$. If 
we consider two slices of the electron bunch and assume that they 
have transverse sizes $\sigma_{x1}$ and $\sigma_{x2}$, then, after 
$N_G$ growth lengths, the relative power in each slice (assuming 
equal initial amplitudes) will satisfy

\begin{equation}
{P_2 \over P_1}=\exp (N_G({\sigma_{x1}^{2/3}\over \sigma_{x2}^{2/3}} -1))
\end{equation}

\begin{figure}[H]
\begin{center}
 \includegraphics[scale=0.30]{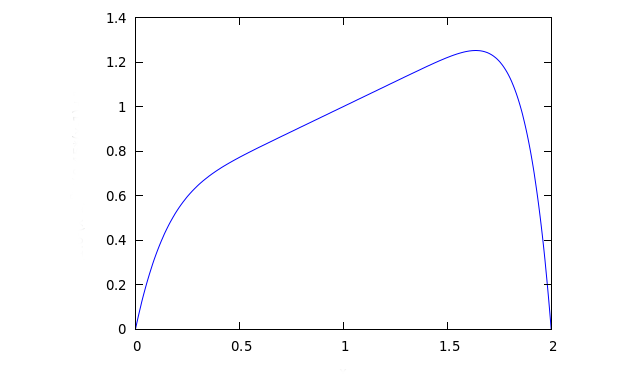} 
\end{center}
\caption{Rough indication of the beam radius needed for the correct variable gain.}
\label{fig:beampro}
\end{figure}

\noindent If we take $\sigma_{x2}=2 \sigma_{x1}$, \ie the forward electrons 
larger transversally than the trailing electrons, then

\begin{equation}
{P_2 \over P_1}=\exp (-0.37 N_G),
\end{equation}

\noindent and after three growth lengths the power should look something 
like figure \ref{fig:initdecrease}. Thus, we need to show how to achieve a transverse 
beam profile that looks roughly like figure \ref{fig:beampro}.

 To achieve such a beam shape requires transverse forces that 
vary from the front to the back of the bunch, \ie time-dependent 
forces. We thus need time dependent electromagnetic fields that 
vary on such a rapid time scale. We give an example of a scheme to do this
in appendix B.

\section{Simulations}

To investigate the efficiency of this scheme we have performed 
a number of simulations of a reduced model of the FEL
with these elements.

We divide the electron beam into a number of slices, each
uniformly about a coherence length in size and divide the photon
beam into a equal number of longitudinal slices. The coherence length,
$L_c$, is given by

\begin{equation}
L_c={\lambda \over 4\pi \rho},
\end{equation}

\noindent where $\lambda$ is the XFEL wavelength, and $\rho$
is the Pierce parameter defined earlier.
In addition,
the electron beam and photon beam are divided into a number
of phases (typically $24$) of the electron waves and photon waves
equally spaced around $360$ degrees.

\subsection{Initialization}
 
\begin{figure}[H]
\begin{center}
\includegraphics[scale=0.350]{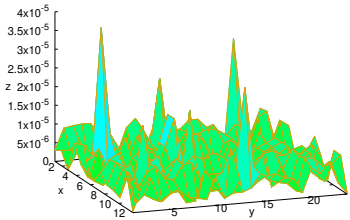} 
\end{center}
\caption{Initial photon beam for one run.}
\label{fig:initone}
\end{figure}

Initially, the electron beam and photon beam are seeded with small
random amplitudes for each slice and each phase. 
For each slice, we then randomly choose one phase and make 
the amplitude for the electron beam and photon beam for that 
phase $4$ times larger. In other words we assume
we have already gone through one section of undulator and that
SASE has produced the typical spiky output at low power. 
The resultant initial beam for one case is shown in
figure \ref{fig:initone}
\noindent As can be seen, this initial beam is not coherent.

\subsection{Time advance}

Once we initialize the beams (electron and photon) we send the
beams through a number of sections where they interact. A section
consists of two parts: 1) an undulator section where the electrons
drive the photons and the photons drive the electrons, and 2) a delay
chicane where the electrons slide back by a slice, i.e. the photons 
slide forward by one slice. We thus describe the electrons as an
array, $e(n,i,j)$ where $n$ labels the time step, $i$ labels the
longitudinal slice number, and $j$ labels the phase number. Likewise,
the photons are described as an array $p(n,i,j)$ where $n$ labels 
the time step, $i$ labels the longitudinal slice number, and $j$ 
labels the phase number. These arrays for $n=1$ (initial time) are
shown in figure \ref{fig:initone} for one case.
 The undulator part of the advance satisfies
the equations

\begin{equation}
e(n+1,i,j)=e(n,i,j)+\Gamma (i)p(n,i,j)
\end{equation}

\begin{equation}
p(n+1,i,j)=p(n,i,j)+\Gamma (i)e(n,i,j)
\end{equation}

 \noindent where $\Gamma (i)$ is a growth that varies along the 
longitudinal slices, $i$.
Note that $\Gamma$ does not depend on the phase; all
phases are assumed to have the same growth rates.

In the chicane part of the advance the photons slip forward
by one slice relative to the electrons while the electrons 
are unchanged (i.e. slip backward relative to the photons). 
Of course the slice of photons in the very front is
simply lost to the simulation (no longer followed), 
and a new slice of photons
with very small random amplitudes for each slice
and phase is injected at the rear. 
The Maxima \footnote{maxima.sourceforge.net} code that performs these advances is shown in
appendix A.

Of course, in the simulation the growth factor $\Gamma (i)$ must
be specified. For a more realistic simulation this would be found by
following the e-beam properties and using the growth formula. In
these simulations we simply specify the growth function assuming
that it comes from an e-beam with radius that increases toward the front.

\subsection{Simulation results}

\begin{figure}[H]
\begin{center}
\includegraphics[scale=0.250]{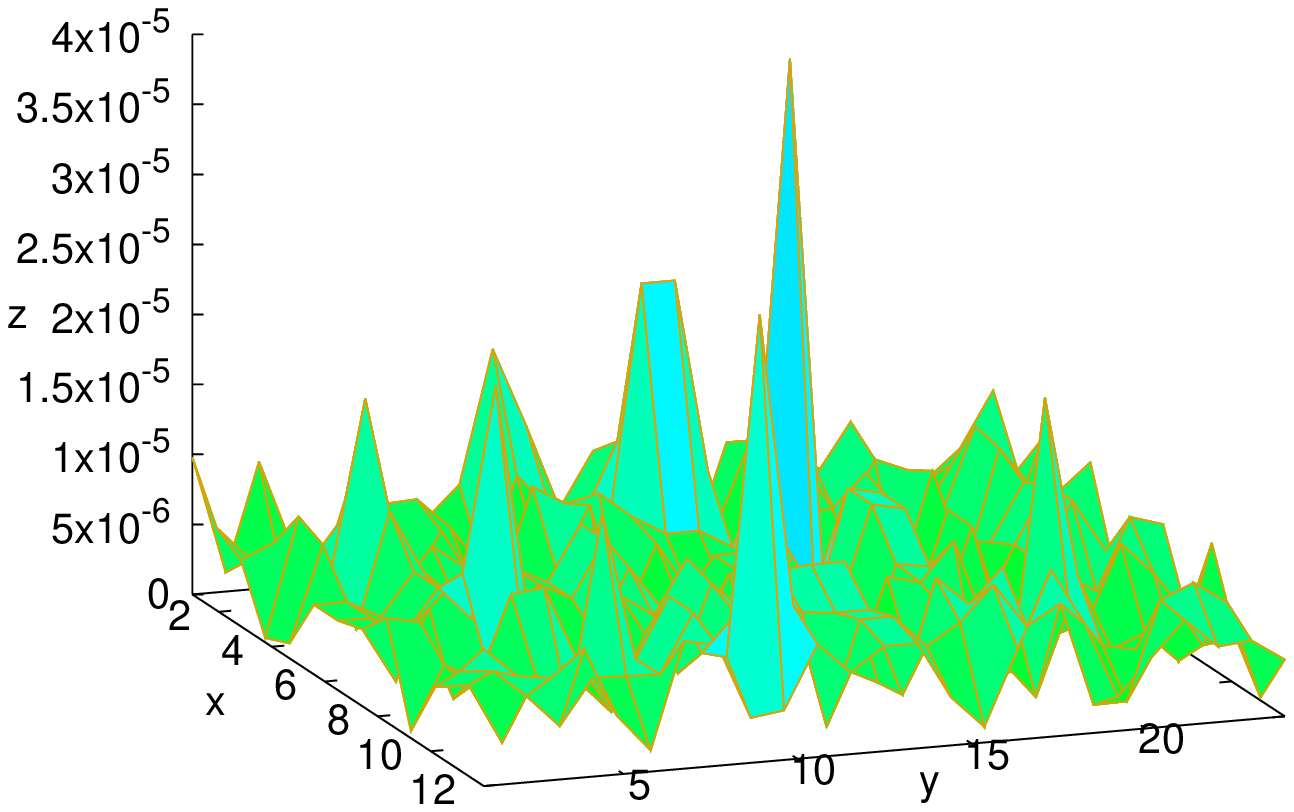} 
\includegraphics[scale=0.250]{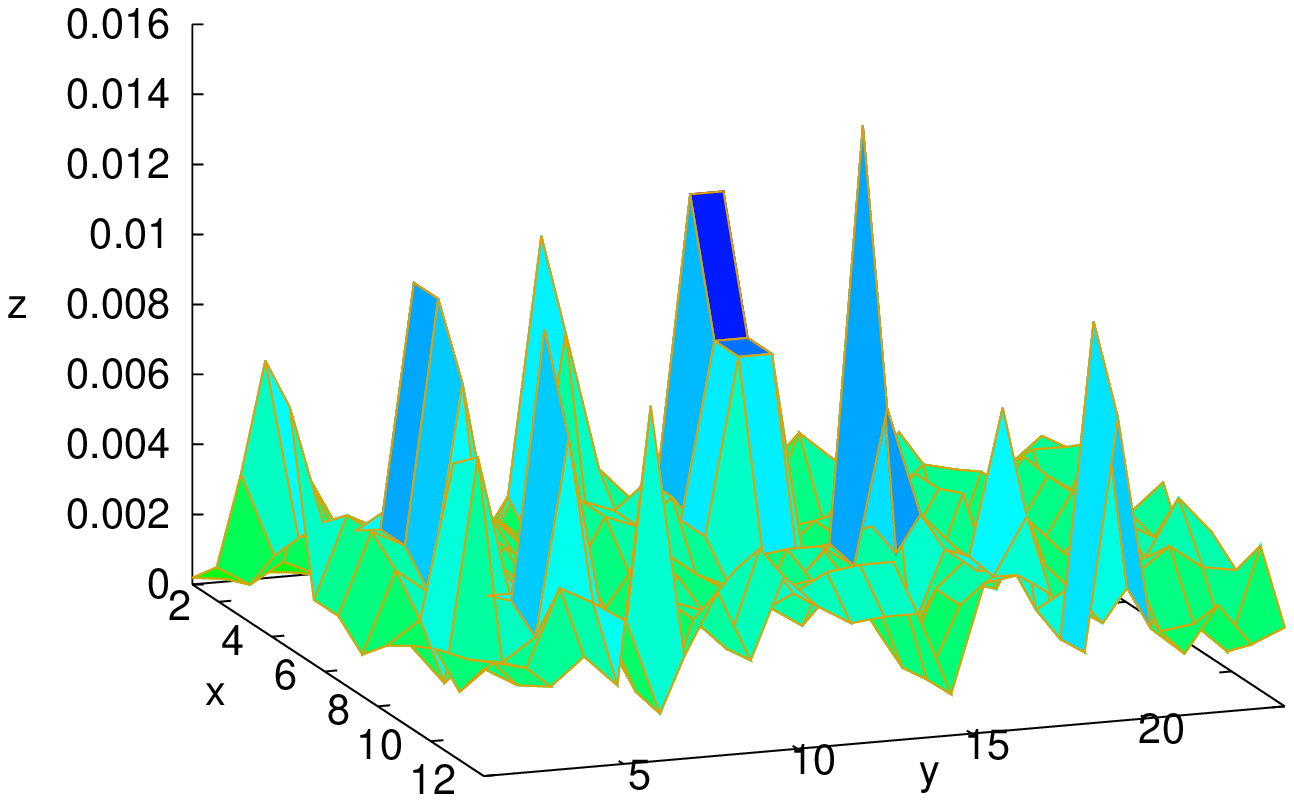}
\includegraphics[scale=0.250]{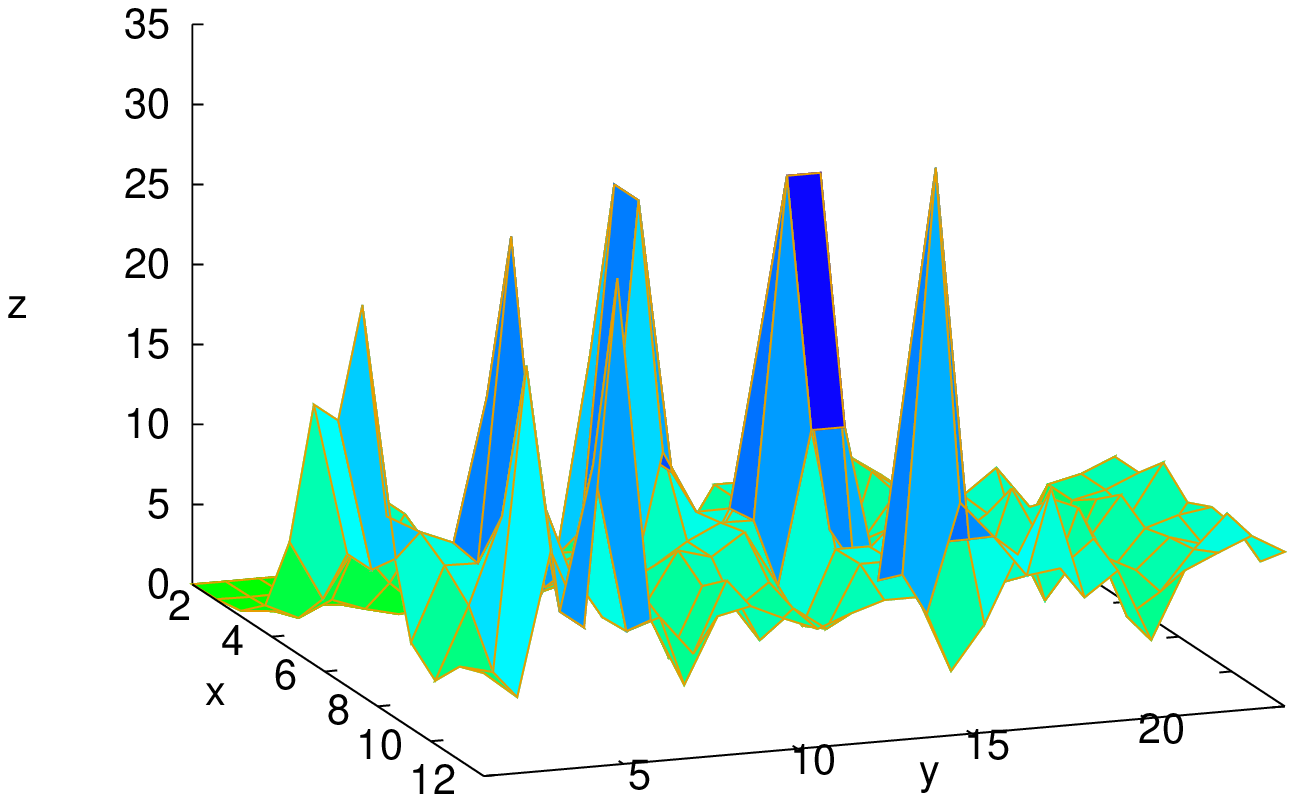}
\end{center}
\caption{For the case with no beam size variation the initial photons (left), after 4 steps (middle), and after 9 steps (right).}
\label{fig:allflat}
\end{figure}

Starting with the initial beams of figure \ref{fig:initone} 
we first tried a simulation with chicanes inserted but with
a constant growth rate for all
phases and all slices, i.e. no beam size variation. The results
at three times are shown in figure \ref{fig:allflat}.

\begin{figure}[H]
\begin{center}
\includegraphics[scale=0.250]{sharp1.eps} 
\includegraphics[scale=0.250]{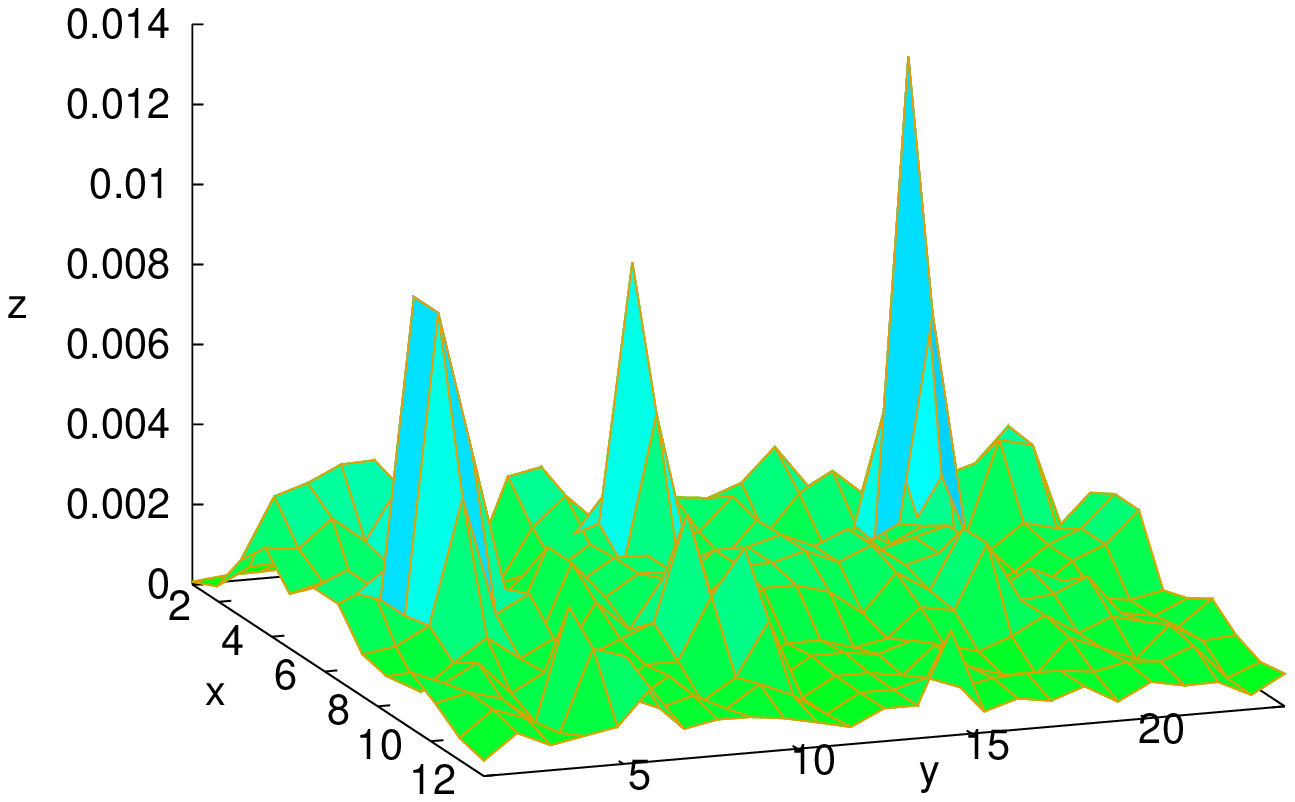}
\includegraphics[scale=0.250]{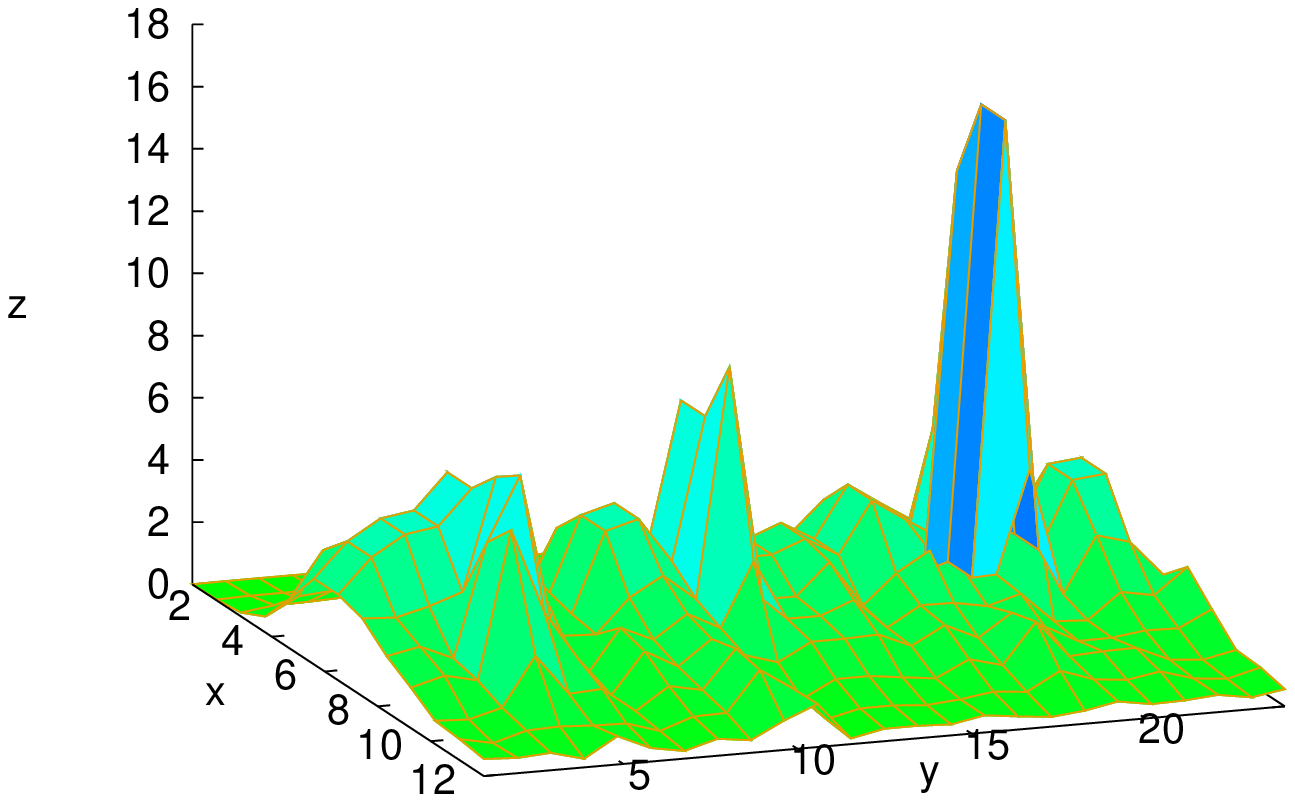}
\end{center}
\caption{For a case with beam size variation the initial photons (left), after 4 steps (middle), and after 9 steps (right).}
\label{fig:decrease}
\end{figure}

\noindent As can be seen there is almost no improvement in the longitudinal coherence.
In a simulation in which the growth rate decreases with the slice number reflecting an
increasing e-beam size toward the front the results are shown in figure \ref{fig:decrease}

\begin{figure}[H]
\begin{center}
\includegraphics[scale=0.50]{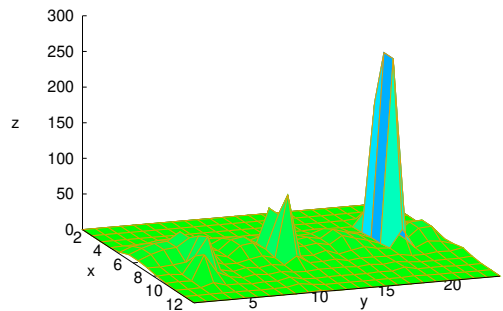} 
\end{center}
\caption{photon beam energy at the final time}
\label{fig:sqdecrease}
\end{figure}

\noindent There is significant improvement in the longitudinal coherence. Also, 
recall that what is shown are amplitudes; the
energies are proportional to the square of the amplitudes
so the energy is almost completely in one phase of the photons.
If we plot the square of the final results of figure \ref{fig:decrease}
we get figure \ref{fig:sqdecrease}

\noindent The concentration on a single phase is evident.
Different values of the constants in the simulation lead to qualitatively
similar results, differing only in the number of steps required to
establish a dominant phase.

\section{Discussion}

We have presented a new scheme for establishing longitudinal coherence
in the output of X-ray FELs. It requires a sequence of undulators and
delay chicanes and the careful tailoring of the initial e-beam with
time dependent focusing. Simulation of a simplified model
of the FEL showed that the idea works extremely well.

Of course, a completely convincing demonstration of this idea would
require more realistic and fully self consistent simulations that would establish
a realistic set of parameters to achieve coherence and determine
the practicality of this scheme.

\section{Appendix A}

In this appendix we present the Maxima simulation code used 
in this paper. The values of the constants are representative
and were varied in different runs.

\begintt
kill(all);
/* # We begin initializing the random number generator. */
/* # Just change nrand for different randoms. */
nrand:17;
for i:1 thru nrand do
    junk:random(1.0);
end;

/* #The number of growth sections is nsteps. */
nsteps:9;
/* # The number of slices in the e-beam and photon beam is nslice. */
nslice:12;
/* # The number of mode phases is nphase. */
nphase:24;

array(tx,nphase);
array(ty,nslice);
array(tstep,nsteps+1);
for i:1 thru nphase do tx[i]:i;
for i:1 thru nslice do ty[i]:i;
for i:1 thru nsteps+1 do tstep[i]:i;

array(p,nsteps+1,nslice,nphase);
array(e,nsteps+1,nslice,nphase);
/* # We initialize the amplitudes of the e-beam phases with small values. */
/* # The first index of e is the step number (starting at 1). */
/* # The second index of e is the slice number. */
/* # The third index of e is the phase number. */
for i:1 thru nslice do  
    (for j:1 thru nphase do  
        e[1,i,j]:0.00001*random(1.0));  

\endtt

\begintt

/* /\* # We initialize the amplitudes of the photon beam phases with small values. *\/ */
/* /\* # The first index of p is the step number (starting at 1). *\/ */
/* /\* # The second index of p is the slice number. *\/ */
/* /\* # The third index of p is the phase number. *\/ */
for i:1 thru nslice do block(  
    for j:1 thru nphase do  
        p[1,i,j]:0.00001*random(1.0));  

/* For each slice we find the max of the random values *?
/* and multiply it by a factor. */

for i:1 thru nslice do block(
  temp[i]:0.0,
  for j:1 thru nphase do block(
    if e[1,i,j] > temp[i] then temp[i]:e[1,i,j]));

for i:1 thru nslice do block(
  for j:1 thru nphase do block(
    if e[1,i,j] = temp[i] then
    block(e[1,i,j]:4.0*e[1,i,j],p[1,i,j]:4.0*p[1,i,j])));

/* # We take the basic growth to be gamma0, giving growth */
/* # of gamma0^2 in a section. */
gamma0:3.9;

/* Choosing alpha=0.5 gives decreasing growth toward the front. */
/* Choosing alpha=0 gives constant growth from front to back. */
 alpha:0.5; 
/*  alpha:0.0; */      
array(slicegrowth,nslice);
/* # We make the growth (slicegrowth) of each slice and each growth step to */
/* # be less than gamma0 depnding on the slice number and step number. */

    for j:1 thru nslice do
        slicegrowth[j]:gamma0*(1-alpha*(j-1)/(nslice));

/* # Each growth step consists of an exponential growth of */
/* # e and p (for all phases) followed by a transfer forward */
/* # of the photon beam (renumbering indices) with a new */
/* # slice of photons injected with a small random value */
/* # for all phases. */
for n:1 thru nsteps do block(
/* # First the growth part. */
    for i:1 thru nslice do block(
        for j:1 thru nphase do block(
            e[n+1,i,j]:e[n,i,j]+slicegrowth[i]*p[n,i,j],
            p[n+1,i,j]:p[n,i,j]+slicegrowth[i]*e[n,i,j])),                         
/* # Store values in a temporary array. */

for i:1 thru nslice do block(
        for j:1 thru nphase do block(
            ptemp[i,j]:p[n+1,i,j])),
/* # Slide the photons forward. */
    for i:2 thru nslice do block(  
        for j:1 thru nphase do block(  
            p[n+1,i,j]:ptemp[i-1,j])),

     for j:1 thru nphase do
         p[n+1,1,j]:0.2*p[n+1,1,j]
   );
\endtt

\begintt
g(i,x,y):=block(x1:round(x),y1:round(y),p[i,x1,y1])$
set_plot_option([elevation,70]);
set_plot_option([azimuth,70]);
set_plot_option([legend,false]);

for istep:1 thru nsteps+1 do
plot3d(g(istep,x,y),[x,1,nslice],[y,1,nphase],
[grid,nslice,nphase],[gnuplot_term,ps],
  [gnuplot_out_file,concat("sharp",istep,".eps")]);

kill(g,x1,y1); 
g(i,x,y):=block(x1:round(x),y1:round(y),p[i,x1,y1]^2)$
set_plot_option([elevation,70]);
set_plot_option([azimuth,70]);
set_plot_option([legend,false]);

for istep:1 thru nsteps+1 do
plot3d(g(istep,x,y),[x,1,nslice],[y,1,nphase],
[grid,nslice,nphase],[gnuplot_term,ps],
  [gnuplot_out_file,concat("sharpsq",istep,".eps")]);

\endtt

\section{Appendix B}

 To achieve such a beam shape requires transverse forces that 
vary from the front to the back of the bunch, \ie time-dependent 
forces. We thus need time dependent electromagnetic fields that 
vary on such a rapid time scale. These forces have to have a 
wavelength about four times the bunch length. For a $20 \mu$m bunch 
we thus need a wavelength of about $100 \mu$m, possibly a laser.
The laser needs to be injected in a structure (possibly a grating)
to make use of the transverse focusing fields (vacuum fields
produce no net focussing). 
Note that the transverse size of the beam should also be a significant
fraction of the wavelength to make efficient use of the field, though 
not so large that nonlinearities become important.

If we assume such forces act over a small distance causing a change 
in transverse angle of $\Delta\theta$ and are followed by a  drift 
of length $L$, then the transverse size change (we assume cylindrical 
symmetry for simplicity) is given by

\begin{equation}
\Delta r=L\Delta\theta,
\end{equation}

\noindent where $\Delta\theta$ and thus $\Delta r$ will vary from particle 
to particle, depending on the forces. Now 

\begin{equation}
\Delta\theta = {\Delta p\over p},
\end{equation}

\noindent where $p$ is the momentum and the change in momentum caused 
by the defocusing forces is

\begin{equation}
\Delta p= eB\Delta t,
\end{equation} 

\noindent with $B$ the magnetic defocusing field, $e$ the charge, 
and $\Delta t$ the time over which the force acts. If we assume 
a set of $N_c$ small cavities, each of length $\Lambda / 2$, then

\begin{equation}
\Delta t= {N_c \Lambda \over 2c},
\end{equation}

\noindent and the angle change is

\begin{equation}
\Delta \theta \approx {eBN_c \Lambda\over 4 p c},
\end{equation}

\noindent where an extra factor of $1/2$ is an approximation of 
the transit time factor. Of course, for relativistic particles, 
$p c \approx {\cal E}$, where ${\cal E}$ is the particle energy.
Thus, requiring $\Delta r=r_B$ for some particles gives the requirement

\begin{equation}
{eBN_c\Lambda \over 4{\cal E}}\approx {r_B\over L}.
\label{condition}
\end{equation}

\noindent If we assume the field extends out to $\approx \Lambda$, 
then the field at the beam edge is

\begin{equation}
B\approx {r_B B_{max}\over \Lambda}.
\end{equation}

\noindent Plugging this into equation \ref{condition} we get 

\begin{equation}
{eB_{max}N_c \over 4{\cal E}}\approx {1\over L}.
\end{equation}

\noindent As an example, if we take

\begin{eqnarray}
{\cal E}=& 12 \, {\rm GeV},\\
N_c=&10 \, ,\\
L=& 5 \, {\rm m},
\end{eqnarray}

\noindent then we find

\begin{equation}
B_{max}\approx 3.3 \, {\rm Tesla},
\end{equation}

\noindent  
a reasonable value at $3.2$ THz. 

\end{document}